# STUDY OF THE VARIABLE STAR ASASSN-V J104912.47+274312.7 A RARE PECULIAR RRc STAR

M. Correa,[1,2] J.M. Vilalta[1,2], J.F. Le Borgne[1,3,4]

1 GEOS (Groupe Européen d'Observation Stellaire), http://geos.upv.es/
2 Agrupació Astronòmica de Sabadell, Prat de la Riba, s/n - 08206 Sabadell (Barcelona), Spain
3 IRAP; OMP; Université de Toulouse; 14, avenue Edouard Belin, F-31400 Toulouse, France.
4 LAM, Laboratoire d'Astrophysique de Marseille, 38 Rue Frédéric Joliot Curie, F-13013 Marseille, France.

## Abstract

During a screening in ASAS-SN database searching candidates of Delta Scuti stars with short period, our attention was drawn to the variable star ASASSN-V J104912.47+274312.7, we considered interesting to follow. It is an ASAS-SN discovery, classified by them as a RRc. We observed it for 16 nights in 2021, obtaining several maxima that allow us to refine its period. After a frequency analysis, we conclude that this star belongs to a small subgroup of RRc stars having the peculiarity to show two close frequencies with ratio greater than 0.96, without 0.61 or 0.68 period ratio mode. This behaviour is very similar to two stars found by Netzel & Smolec (2019) in OGLE data, and one star in NGC 6362, studied by Smolec et al. (2017). This large period ratios are also found as additional mode to standard double mode RRc stars ($P_x/P_l$= 0.61-0.68) by Netzel et al. (2023).

## 1 Observations

Mercè Correa has observed the field of ASASSN-V J104912.47+274312.7 during eleven nights with a Meade 12" f/5.8 telescope and ccd camera through Johnson V filter from her observatory in Freixinet (Catalonia, Europe). Matteo Monelli has observed the star during 2 nights from Teide Observatory (Canary Islands), with the telescope IAC80 with an effective focal ratio of f/11.3, effective focal length of 9.02 m and a primary mirror of diameter 82 cm, equipped with Camelot2 CCD. Details of their observations are given in Tab. 1. The comparison (C1) and check (C2) stars used are given in Tab. 2 together with coordinates and magnitudes from GSC2.3 catalog (Lasker et al. 2008). A view of the field is shown in Fig.1.

| Observer | Telescope | Filter | # Nights | # meas. | First night JD | Last night JD |
|---|---|---|---|---|---|---|
| M. Correa | Meade 12" | V | 11 | 3619 | 2459289.34375 (2021-03-15) | 2459322.70139 (2021-04-18) |
| M. Monelli | IAC 80 | V | 2 | 146 | 2459330.33333 (2021-04-25) | 2459332.60625 (2021-04-28) |

Table 1. Observations of ASASSN-V J104912.47+274312.7

|  | GSC2.3 | RA | DEC | F (R phot) | B (phot) | V | B-V |
|---|---|---|---|---|---|---|---|
| ASASSN-V J104912.47+274312.7 | N6M5000044 | 162.302008 | 27.7202 | 12.75 | 13.18 | 12.98 | 0.20 |
| C1 | N6M5000035 | 162.263345 | 27.794777 | 12.32 | 13.12 | 12.68 | 0.44 |
| C2 | N6M5000041 | 162.266637 | 27.742113 | 12.53 | 13.47 | 12.94 | 0.53 |

Table 2. Coordinates and magnitudes of ASASSN-V J104912.47+274312.7 and comparison stars from GSC2.3 catalogue

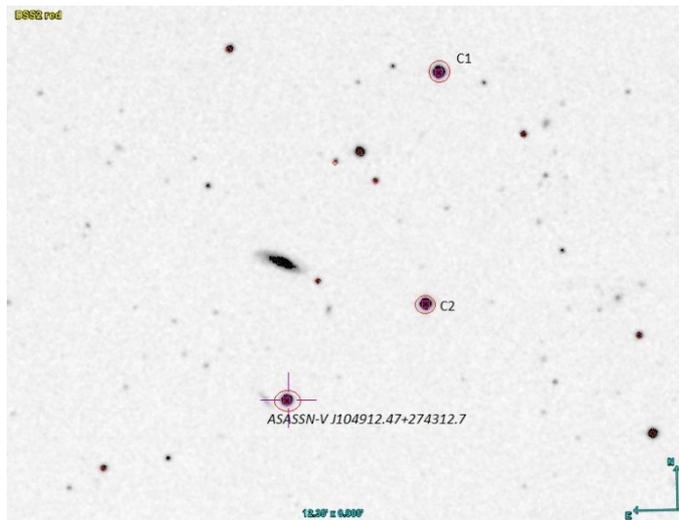

*Figure 1. Finding chart of the field of ASASSN-V J104912.47+274312.7 (id. in GSC2.3 N6M5000044)*

## 2　Data reduction

The data reduction including image calibration and photometric measurements were performed by Mercè Correa with software AIP4WIN (Berry & Burnell, 2005). The calculation of maxima was done with a software based on splines, developed by Josep M. Vilalta. The method consists in smoothing the light curve by series of spline functions following Reinsch's algorithm (Reinsch, 1967).

On the first night of observation, two consecutive maxima were detected with an O-C of 0.0229 with respect to the ephemeris calculated with the ASAS-SN elements. Which led us to think that the period might require fine-tuning. The successive observations showed a highly fluctuating O-C (see Tab. 3). The elements of ASAS-SN were at that time: 2458132.97043 + 0.2659 E.

Matteo Monelli recalculated the period from ASAS-SN observations proposing a period of 0.2623803 d. Taking ASAS-SN HJD 2458132.97043 as the epoch, and the new Monelli's period of 0.2623803 days, the ephemeris and the corresponding O-C were calculated, which are shown in Tab. 3.

| HJD | Unc. (Day) | O - C (Day) | Observer |
|---|---|---|---|
| 2459289.38701 | 0.0049 | 0.107 | M. Correa (AAS) |
| 2459289.64929 | 0.0045 | 0.106 | M. Correa (AAS) |
| 2459294.45806 | 0.0076 | -0.070 | M. Correa (AAS) |
| 2459296.54727 | 0.0054 | -0.080 | M. Correa (AAS) |
| 2459302.40063 | 0.0056 | 0.001 | M. Correa (AAS) |
| 2459303.44736 | 0.0066 | -0.002 | M. Correa (AAS) |
| 2459305.34370 | 0.0062 | 0.058 | M. Correa (AAS) |
| 2459306.43029 | 0.0068 | 0.095 | M. Correa (AAS) |
| 2459310.37052 | 0.0059 | 0.100 | M. Correa (AAS) |
| 2459323.42573 | 0.0076 | 0.036 | M. Correa (AAS) |
| 2459332.45041 | 0.0034 | -0.123 | M. Monelli (IAC) |

*Table 3. Relation of observed maxima and their corresponding O-C*

## 3  Period variation and O-C modulation

A frequency analysis was carried out on the basis of the observations of ASASSN J104912.47+274312.7 until April 27, 2021. Using Period04, 3740 points were processed, obtaining the frequencies:

|    | c/d         | amplitude   | phase       | snr   |
|----|-------------|-------------|-------------|-------|
| f0 | 3.760854534 | 0.09128077  | 0.12533634  | 152.6 |
| f1 | 3.926499311 | 0.05046696  | 0.60477926  | 86.5  |
| f2 | 0.032560503 | 0.01016672  | 0.62206268  | 12.1  |
| f3 | 7.520493591 | 0.00734914  | 0.94936243  | 13.8  |
| f4 | 1.012833219 | 0.01030217  | 0.19623352  | 12.7  |
| f5 | 7.841715704 | 0.00489271  | 0.52931035  | 9.6   |
| f6 | 2.708872106 | 0.0040448   | 0.49936636  | 5.9   |
| f7 | 3.97586915  | 0.00315127  | 0.72243661  | 5.4   |
| f8 | 9.202521486 | 0.00193466  | 0.2583889   | 4.1   |
| f9 | 7.430578638 | 0.00206011  | 0.14379373  | 4.0   |

Fundamental period f0 = 3.760854 c/d (0.265897 d)

Figure 2 shows some observed points and the fitted trigonometric function and Fig. 3 shows the phase diagram based on a period corresponding to f0 = 3.760854534 c/d (0.265897 d).

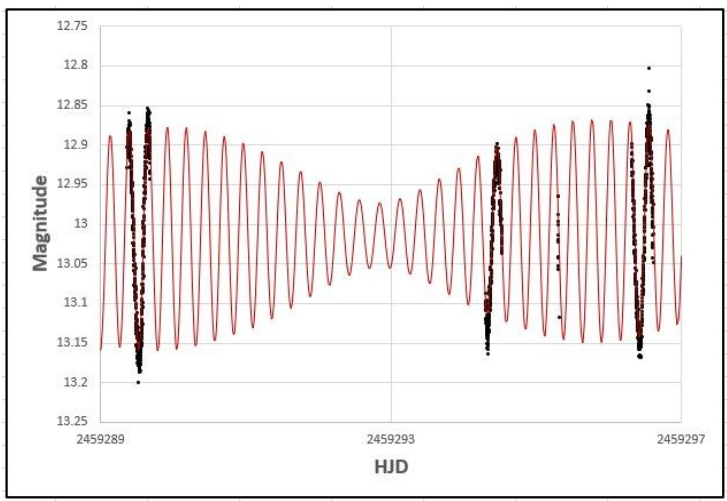

Figure 2. Partial view of the adjustment of our observations.

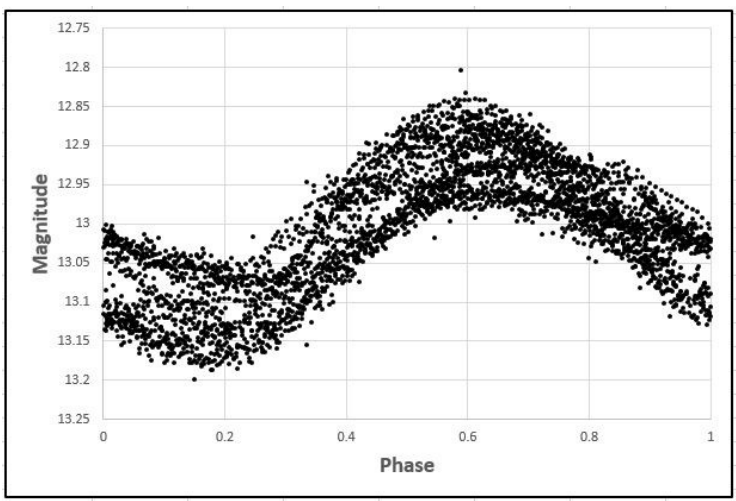

Figure3. Phase diagram

The phased light curve, presented in the Fig. 3, shows that the star is multiperiodic or modulated. A periodic variation is also found in the O-C of observed maxima. Examining the list of extracted frequencies, we find the frequency f1 (3.926499 c/d) quite close to the fundamental frequency f0 (3.760854534 c/d). The difference of the two frequencies, f1–f0= 0.1656 c/d, corresponds to a modulation period of 6.0386 days.

The O-C of maxima obtained by M. Correa, phased on the above modulation period, are plotted in Fig. 4. The epoch is arbitrarily taken on JD 2459302.9. The O-Cs are calculated with the period 1/f0=0.265897 days which eventually better accounts for the observations than Monelli's period.

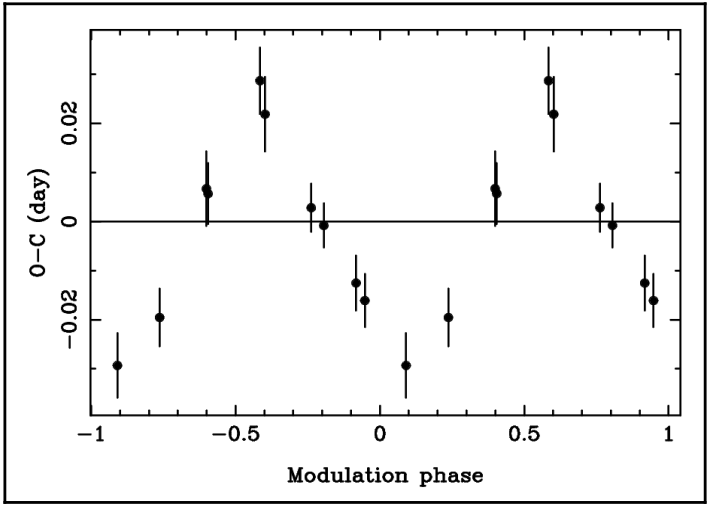

*Figure 4. O-C of maxima obtained by M. Correa, phased on the modulation period.*

To enhance the analysis of this star, photometric 897 data points from TESS (Transiting Exoplanet Survey Satellite, Ricker et al., 2015) were processed, obtaining the following frequencies which are similar to those found from our data:

| | |
|---|---|
| f0 = 3.760749 | f3 = 7.853028   (2f1) |
| f1 = 3.926883 | f4 = 0.184044 ≈(f1-f0) |
| f2 = 7.521250   (2f0) | f5 = 7.687490   (f0+f1) |

Figure 5 shows the Fourier fit of TESS data:

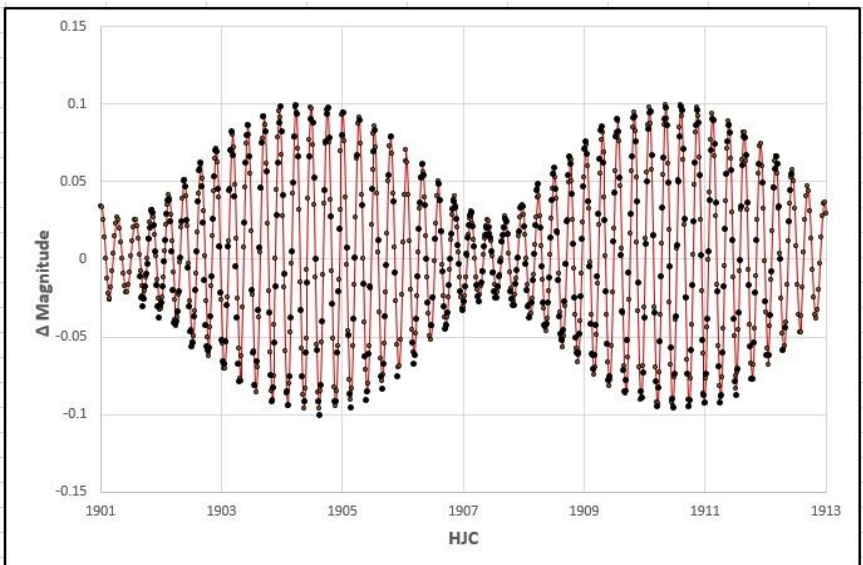

*Figure 5. Fourier adjustment of the TESS observations.*

## 4   Discussion

Netzel & Smolec (2019) analyzed the photometry of a group of first-overtone RR Lyrae stars (RRc; 11415 stars) and double-mode RR Lyrae stars (RRd; 148 stars) towards the Galactic bulge from the Optical Gravitational Lensing Experiment (OGLE: Udalski et al. 1997).

They have also detected two double-periodic RRc variable stars, with two close periodicities, similar to RR Lyr variable V37 in NGC 6362 analyzed by Smolec et al. (2017).

They consider beating to be the interaction of the two close frequencies instead of amplitude modulation. The period ratio P1/P0 is 0.98-0.99 while for a typical RRc it is 0.61 or 0.68 and for RRd it is about 0.75 (Netzel & Smolec, 2019; Netzel et al., 2023). These peculiar variables could form a new sub-group of double-mode pulsators.

We have located the variable under study ASASSN-V J104912.47+274312.7, on the Period-Amplitude diagram for RRc stars published by Netzel & Smolec (2019), where they located their two anomalous variables and the previous found V37 (NGC 6362), see Fig. 6. All of them are located in the short period and low amplitude region.

The period ratio of ASASSN-V J104912.47+274312.7 is P1/P0 = 0.9578

It seems therefore that the variable star ASASSN-V J104912.47+274312.7 has similar characteristics to those studied by Smolec et al. (2017) and Netzel & Smolec (2019) in the above-mentioned papers.

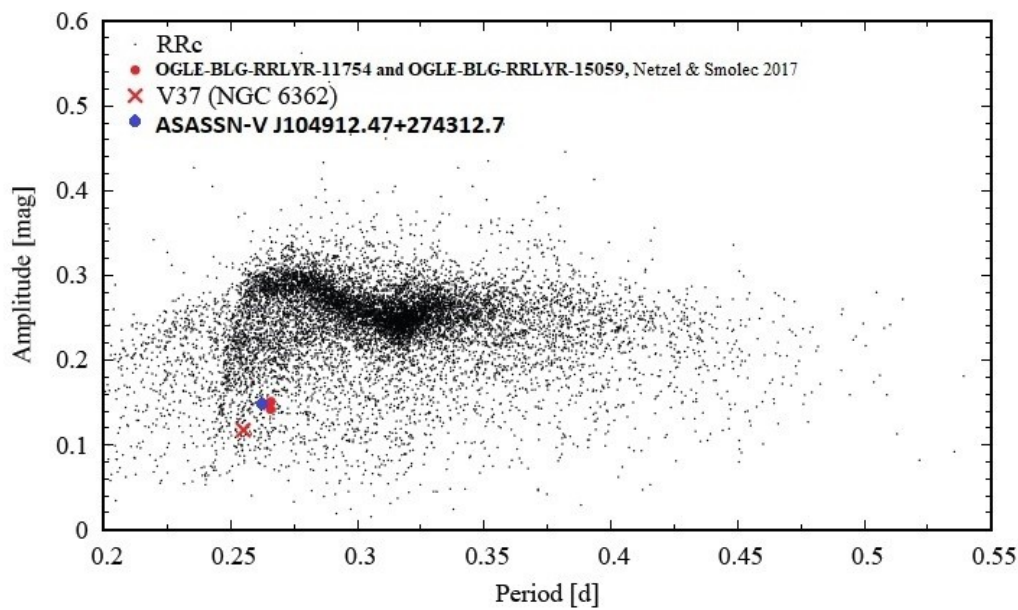

*Figure 6. Period-amplitude diagram for RRc stars (black dots), OGLE-BLG-RRLYR-11754 and OGLE-BLG-RRLYR-15059 (red point), V37 star in NGC6362 (red cross) and ASASSN-V J104912.47+274312.7 (blue circle). Modified from Fig. 14 in Netzel & Smolec (2019).*

The two periodicities were separated in TESS data as shown in Fig. 7, which displays in the top panel the light curve phased with P0 = 0.265904 d, the large scatter shows the beating of the two periods: P0 and the close one P1 = 0.254654 d.

In the middle and the bottom panels are represented the disentangled light curves for the periods P0 and P1. These curves are very symmetrical and almost sinusoidal. In contrast, the curve of the first overtone presented by Smolec et al. (2017) is quite asymmetrical.

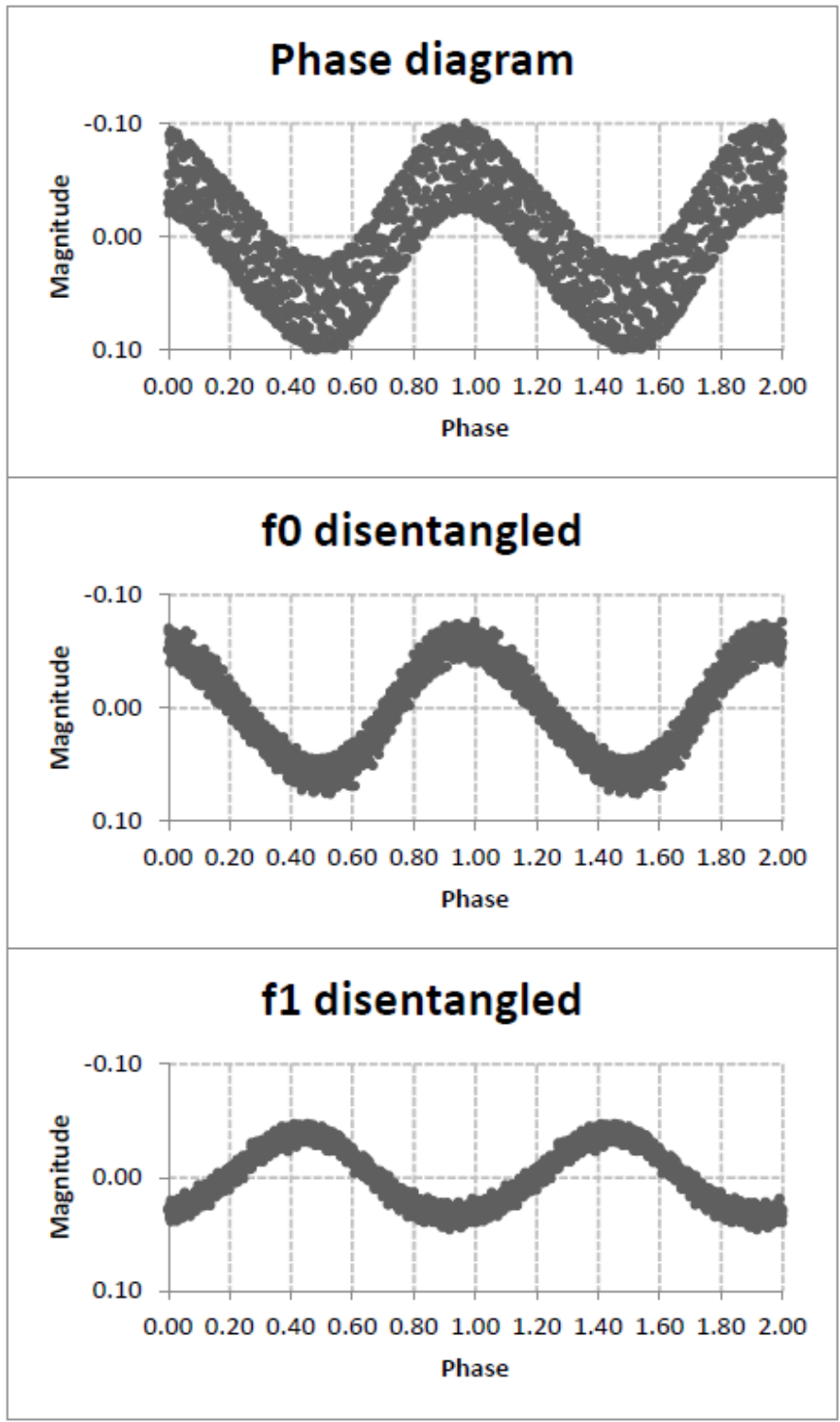

Figure 7. Light curves of TESS observations for ASASSN-V J104912.47+274312.7. Top panel the light curve phased with P0 = 0.265904 d. Middle and the bottom panels with the disentangled light curves for the periods P0 and P1.

Table 4 summarizes the comparison of the frequency analysis of the three data sources analysed, our data, ASAS-SN (V and g filters) (Shappee et al. 2014, Kochanek et al. 2017) and TESS. At time of download (July 2023), ASAS-SN database contained 869 measurements in V filter obtained during 334 nights between JD 2455938 and 2458448 and, for g filter, 2107 measurements during 639 nights (JD 2458063-2460133). The columns contain the values of first-overtone frequency f0, the associated period P0, the additional frequency f1, the frequency ratio f0/f1 and frequency difference f1-f0.

|            | f0 (c/d)  | P0 (d)   | f1 (c/d)  | f0/f1   | f1-f0 (c/d) |
|------------|-----------|----------|-----------|---------|-------------|
| Our data   | 3.760855  | 0.265897 | 3.926499  | 0.95781 | 0.1656      |
| ASAS-SN V  | 3.760730  | 0.265906 | 3.926730  | 0.95772 | 0.1660      |
| ASAS-SN g  | 3.760760  | 0.265904 | 3.926760  | 0.95772 | 0.1660      |
| TESS data  | 3.760749  | 0.265904 | 3.926883  | 0.95769 | 0.1661      |

*Table 4. Comparison of the three data sources analysed (ASAS-SN, Tess and our data).*

Recently, Netzel et al. (2023) published an analysis of first-overtone RR Lyrae stars from Kepler satellite K2 data. While most of the double mode stars they study fall in the Petersen diagram in the sequences with period ratios of 0.61 and 0.68, they also find additional signals in a third frequency. The stars with this additional signal the origin of which is not known populate the Petersen diagram with period ratio ~0.5, ~0.75, ~0.85 and above 0.94 (See Fig. 18 in Netzel et al., 2023). *ASASSN-V J104912.47+274312.7 belongs to* this last sequence. However, of the stars studied by Netzel et al. (2023), few of them (9 in the ~0.85 Ps/Pl sequence) show these signals and not the 0.61 period ratio.

In GAIA satellite DR3 variability classification (Gaia Collaboration, 2022b), ASASSN-V J104912.47+274312.7 (source id. 730980520129043072) appears as a RRc variable star with a period of 0.265913 days but it fails to find a second frequency because of the small number of measurements.

Concerning the distance and the physical parameters of the star, GAIA DR3 and EDR3 derived publications give some information, sometime contradictory. Bailer-Jones et al. (2021) give a distance of 3734.589 pc while Anders et al. (2022) and GAIA collaboration (2022a) give more coherent values of 2980.115 pc and 2787.9016 pc respectively.

In table 5 we summarise the physical parameters as given by Anders et al. (2022) and GAIA collaboration (2022a). There are discrepancies between these values thought none of them are incompatible with what is expected for a RR Lyr star, except the mass given by GAIA collaboration (2022a) (FLAME, Creevey et al. 2023) which is about four times too large. However, in GAIA documentation[1] it is said that FLAME parameters estimation is based on solar metallicity models, and should not be used for low metallicity stars like ASASSN-V J104912.47+274312.7. We should then trust Anders et al. (2022) values rather than those of GAIA collaboration (2022a).

|                      | Mass ($M_\odot$) | Teff (K) | log(g) (cm/s2) | [Fe/H]    | Abs. G mag. | BP-RP$_0$ |
|----------------------|---------|----------|----------------|-----------|-------------|-----------|
| Anders et al. (2022) | 0.789   | 7518.44  | 3.057061       | -2.167642 | 0.368969    | 0.264555  |
| GAIA coll. (2022)    | 3.146   | 9652.3   | 3.7309         | -0.5925   | 0.0206      |           |

*Table 5. Physical parameters from GAIA DR3 as published by Anders et al. (2022) and GAIA collaboration (2022) .*

If we refer to figure 27 of Netzel et al. (2023), the absolute magnitude and color index place the star on the blue edge of their sample of first-overtone RR Lyr stars. We note their comment on the fact that pure overtone pulsators are mainly on the blue edge of the color-magnitude diagram while stars with extra modes (Px/P1o = 0.61 or 0.68) are cooler and lie toward the red edge. Since ASASSN-V J104912.47+274312.7 has no Px/P1o = 0.61 or 0.68 pulsation mode, it then has to be associated with pure overtone pulsators, the additional

---
1   https://gea.esac.esa.int/archive/documentation/GDR3/

mode close to first-overtone being a phenomenon radically different from the classical extra modes.

Note that GAIA collaboration (2022a) give a rather high proper motion velocity of 24.02 km/s which, given the low metallicity from table 5, indicates that this star probably belongs to the halo or thick disk of the Galaxy.

## 5 Conclusions

We show that the RRc star, ASASSN-V J104912.47+274312.7, is similar to two stars found by Netzel & Smolec (2019): OGLE-BLG-RRLYR-11754 and OGLE-BLG-RRLYR-15059, and V37 star in NGC6362, studied by Smolec et al. (2017), in which beating of two close frequencies is detected.

This peculiar variable has a period ratio P0/P1 of 0.96 out of the normal range of an RRc star (0.61 or 0.68). The resulting light curve is the combination of two symmetrical light curves that are not typical of a RRc. It belongs to a new subclass of RRc stars proposed by Netzel & Smolec (2019). Such a period ratio has also been detected by Netzel et al. (2023) in RRc stars from Kepler K2 sample but as a third additional pulsating mode of stars showing double pulsating mode with a classical period ratio of 0.61 or 0.68.

## 6 Acknowledgements

We thank E. Poretti (GEOS) for useful discussions and M. Monelli (IAC) for collecting the measurements with the IAC80 telescope.

This paper includes data collected by the TESS mission. Funding for the TESS mission is provided by the NASA Science Mission Directorate.

The present study makes use of the following facilities:
- The GEOS database of RR Lyr stars hosted at Institut de Recherche en Astrophysique et Planétologie, Toulouse, France http://rr-lyr.irap.omp.eu/dbrr/ (Le Borgne et al., 2007).
- The SIMBAD database, operated at CDS, Strasbourg,France (Wenger et al., 2000) and the VizieR catalogue access tool also at CDS. The original description of the VizieR service was published in A&A, Supp. 143, 23, http://vizier.u-strasbg.fr/viz-bin/VizieR and of "Aladin sky atlas" developed at CDS, Strasbourg Observatory, France (Bonnarel et al., 2000; Boch & Fernique, 2014).
- The International Variable Star Index (VSX) database, operated at AAVSO, Cambridge, Massachusetts, USA